# The GWR route map: a guide to the informed application of Geographically Weighted Regression

Alexis Comber[1*], Chris Brunsdon[2], Martin Charlton[2], Guanpeng Dong[3], Rich Harris[4], Binbin Lu[5*], Yihe Lü[6], Daisuke Murakami[7], Tomoki Nakaya[8], Yunqiang Wang[9], Paul Harris[10]

[1] School of Geography, University of Leeds, Leeds, UK.
[2] National Centre for Geocomputation, Maynooth University, Maynooth, Ireland.
[3] School of Environmental Sciences, University of Liverpool, Liverpool, UK.
[4] School of Geographical Sciences, University of Bristol, Bristol, UK.
[5] School of Remote Sensing and Information Engineering, Wuhan University, Wuhan, China.
[6] State Key Laboratory of Urban and Regional Ecology, Research Center for Eco-Environmental Sciences, Chinese Academy of Sciences; Joint Center for Global Change Studies; University of Chinese Academy of Sciences, Beijing, China.
[7] Department of Data Science, Institute of Statistical Mathematics, Tachikawa, Japan.
[8] Graduate School of Environmental Studies, Tohoku University, Sendai, Japan.
[9] State Key Laboratory of Loess and Quaternary Geology, Institute of Earth Environment, Chinese Academy of Sciences, Xi'an, China.
[10] Sustainable Agriculture Sciences North Wyke, Rothamsted Research, Okehampton, UK.

[*] corresponding authors: a.comber@leeds.ac.uk binbinlu@whu.edu.cn

## Abstract

*Geographically Weighted Regression (GWR) is increasingly used in spatial analyses of social and environmental data. It allows spatial heterogeneities in processes and relationships to be investigated through a series of local regression models rather than a global one. Standard GWR assumes that the relationships between the response and predictor variables operate at the same spatial scale, which is frequently not the case. To address this, several GWR variants have been proposed. This paper describes a route map to inform the choice of whether to use a GWR model or not, and if so which of three core variants to apply: a standard GWR, a mixed GWR or a multiscale GWR (MS-GWR). The route map comprises primary steps: a basic linear regression, a MS-GWR, and investigations of the results of these. The paper provides guidance for deciding whether to use a GWR approach, and if so for determining the appropriate GWR variant. It describes the importance of investigating a number of secondary issues at global and local scales including collinearity, the influence of outliers, and dependent error terms. Code and data for the case study used to illustrate the route map are provided, and further considerations are described in an extensive Appendix.*

## 1. Introduction

This paper provides guidance for reliable application of Geographically Weighted Regression (GWR). Its aim is to ensure that the increasing numbers of GWR applications in the physical and environmental sciences are correctly formulated and appropriate to the study objective because many are not, even in the peer reviewed literature. GWR is a spatially varying coefficient (SVC) model that quantifies variations in the scales of processes and the relationships being examined. This provides an advantage over alternatives which commonly assume data relationships are fixed (i.e. constant across space), including those accounting for spatial autocorrelation effects. GWR supports enhanced understanding of geographical processes through the study of relationship heterogeneity, which can be a study aim in itself or used to guide further data collection and analysis.

### 1.1 GWR in context

There is a long history of explicitly spatial analyses in a number of disciplines including crop science (Fisher 1935), meteorology (Kolmogorov 1941), geology/mining (Matheron 1963), forestry (Matern 1960), ecology (Clark and Evans 1954), soil science (Burgess and Webster 1980a; b) and geography (Chorley and Haggett 1967). Most developments have centred on accounting for spatial autocorrelation effects (e.g. Cressie 1993) rather than spatial heterogeneity effects (Fotheringham et al. 2002; LeSage and Pace 2009), the latter of which are relatively recent when specifically considering the nature of data relationships in regression models.

GWR (Brunsdon et al. 1996; 1998a) investigates how and if relationships between response and predictor variables vary across space. It is underpinned by the idea that *whole map* regressions such as those estimated by ordinary least squares (OLS) may make unreasonable assumptions about the stationarity of the regression coefficients under investigation (Openshaw 1996; Fotheringham and Brunsdon 1999). As an SVC model, GWR provides measures of non-stationarity in data relationships through the generation of mappable regression coefficients, and inferences on stationarity through statistics and simulation tests (e.g. Nakaya 2015; da Silva and Fotheringham 2016; Harris et al. 2017). As described in Brunsdon et al. (1996; 1998a), GWR stems from locally weighted regression (LWR) (Cleveland 1979; Loader 2004) and thus extensively borrows from the same non-parametric regression paradigm (Wand and Jones 1995), including generalized additive models (GAMs) (Hastie and Tibshirani 1986). As with LWR, GWR is a localized, non-stationary adaptation of the basic linear regression model, where for LWR *localness* is in attribute-space, whilst for GWR *localness* is in geographic-space (see also Páez et al. 2011).

GWR is not a unique concept for SVC modelling. SVC models that pre-date GWR include the expansion method (Casetti 1972) and weighted spatial adaptive filtering (Gorr and Olligschlaeger 1994). Since GWR circa 1996, alternative SVC models have been developed including a parametric version of GWR (Páez et al. 2002a; b), Bayesian SVC models (Assunção 2003; Gelfand et al. 2003), spatial additive models (e.g., Fahrmeir et al., 2000, 2004), and eigenvector spatial filtering (ESF) (Griffith 2003; 2008). Theory for GWR (in its usual non-parametric form), Bayesian SVC, spatial additive, and ESF models has continued to evolve (e.g. GWR models – Brunsdon et al. 1998b; 1999; 2012; Fotheringham et al. 2002; Wheeler 2009; Mei et al., 2016; Geniaux and Martinetti 2018; Yu et al. 2019; e.g. Bayesian SVC models – Waller et al. 2007; Wheeler and Waller 2009; Finley 2011; Datta et al., 2016; e.g. spatial additive models – Kneib et al., 2009; Xue and Liand 2010; e.g. ESF models – Griffith 2012; Murakami et al. 2017; Murakami and Griffith 2019a; b), including many useful SVC model comparison studies using simulated data with known and testable properties (Finley 2011; Oshan and Fotheringham 2018; Wolf et al.

2018; Murakami et al. 2019), demonstrating both the merits and drawbacks of each method. Of these, GWR has been extensively applied in a wide variety of scientific disciplines, such as environment health (e.g. Yoneoka et al. 2016), landscape ecology (e.g. Zhang et al. 2004), soil quality (e.g. Song et al. 2016), air quality (e.g. You et al. 2015), water quality (e.g. Sun et al. 2014), remote sensing (e.g. Foody 2003; 2004), disease patterns (e.g. Brunton et al. 2017), urban studies (e.g. Huang et al. 2019) and housing markets (e.g. Yu et al. 2007).

## 1.2 Motivation: Why this paper is appropriate now?

The motivation for this paper at this time is because GWR is increasingly being used for different spatial analyses but not always correctly. A search of Web of Science (http://apps.webofknowledge.com) for the keyword *Geographically Weighted Regression* in July 2019 indicated 1795 records, with sharp increases in recent years, most articles from USA and China. This proliferation has been driven by a number of factors.

First is the increasingly spatial nature of data, which are now routinely collected with location attached, facilitated by the many GPS-enabled monitoring devices and the tagging of, for example, administrative data with census geographies. Second, there is a broader cross-disciplinary demand for methods to quantify spatial patterns in data, commonly through some kind of hotspot estimation, spatial cluster analysis or spatially informed regression technique. This has been accompanied by recognition of the need to cater for spatial dependencies in the data or the model parameters themselves, reflecting Tobler's first law of geography (Tobler 1970) which describes spatial dependency and spatial autocorrelation. GWR is a method that enables this, and on first sight it appears relatively intuitive model to understand. Third, GWR's simplicity fuels its popularity, which is reflected by its implementation in a number of software packages including the ESRI ArcGIS suite of tools, five R packages (*spgwr* (Bivand et al. 2013), *gwrr* (Wheeler 2013), *GWmodel* (Lu et al. 2014; Gollini et al. 2015), *McSpatial* (McMillen 2013) and *lctools* (Kalogirou 2019)), two Python packages (*PySal* (Rey and Anselin 2010) and *mgwr* (Oshan et al. 2019)) and standalone implementations such as *GWR3* (Charlton et al. 2003), *GWR4* (Nakaya 2015) and *MGWR 1.0* (Li et al. 2019). Each software package has a standard GWR option complemented by a variety of alternative GWR forms and associated tools. No single package provides a fully comprehensive choice to the user although the *GWmodel* package comes closest.

Consequently, it is increasingly easy to find applications of GWR in the literature where it is questionable whether the authors fully understood the inputs, the model assumptions, the model outputs and the associated limitations of different parameter and model choices. The situation is analogous to the old joke *What is a lecture?*[1], and the result is GWR applications that are inappropriate (i.e. where GWR should not have been applied to the problem), poorly calibrated (i.e. the GWR model is incorrectly parameterised), that use the incorrect form of GWR or where the GWR analysis is partial and incomplete. With that in mind, this paper aims to provide a route map to promote the informed use and application of GWR.

## 1.3 A GWR route map

The GWR route map is achieved empirically through a soil case study in the Loess Plateau of China, that guides the reader through different modelling scenarios that are of *primary* importance to a GWR analysis. These main arteries of the route map take the reader to *GWR Basecamp*.

---

[1] Answer: A lecture is the process by which the lecturer's words, as presented on a blackboard, whiteboard or screen, are transcribed to the student's notes *without going through the brain of either*.

Strategies for *secondary* model decisions (*scaling the summit*) are described in the Discussion (Section 4) that outlines a number of *secondary* issues and considerations. Not all *secondary* issues may appear in a specific GWR analysis, and some may interact, including interactions with those considered of *primary* importance. Thus, although the GWR route map is presented as a linear workflow, it should be recognised that in practice, it is often an iterative, more complex process, as may be the case in any regression study. The implications in this respect are that, for some spatial processes, a GWR analysis can be relatively straight-forward, while for others, it can be problematic with increasing complexity. Ultimately, the result of this two-stage *primary-to-secondary* strategy should lead to an informed, sensible and appropriate GWR implementation, from which reliable and robust inferences can be made.

Further, the intention of this paper is not necessarily to replace existing guides, such as high-quality user manuals provided with many of the listed software packages above, but instead to provide a guide that is complementary. This paper also aims to update best practise in GWR modelling, say for example, with respect to the accessible classic GWR text of Fotheringham et al. (2002). Although the topic of this paper sits within the general category of model selection in regression (e.g. Fox 2016) and in spatial regression (e.g. Anselin 2006), its objectives are not to advance the theory in the respect, but instead utilise known theory within a GWR context.

Some generic considerations and further guidance are also given in an Appendix with respect to: (i) sample and data characteristics (Appendix section A1), (ii) influences on weighting schemes (Appendix A2), (iii) inference in GWR (Appendix A3), (iv) GWR as a spatial predictor (Appendix A4) and (v) GWR development through simulation experiments (Appendix A5). The Appendix is extensive covering many important issues and should not be over-looked. In this respect, early drafts of this paper considered many such issues within the main text but were ultimately consigned to the Appendix for narrative purposes.

The route map is presented using only real data. This is deliberate, as the intention is to provide 'real world' practical guidance to a GWR analysis. A statistically rigorous evaluation of the proposed route map through a Monte Carlo simulation experiment that generates data with known properties would be a more involved study, best presented elsewhere and to a different audience. That said, Appendix A5 provides guidance to the implementation of such a study.

## 1.4 Context and analogous extensions

This paper focuses on GWR applications in the physical and environmental sciences where data are commonly measured on a point support. The main messages of the paper are similarly relevant to applications using data measured on area support, for example socio-economic studies of population demographics, inequalities, education, crime or health, as the different implementations of GWR use the areal unit centroids, and thus default to data on point support. Notable exceptions are highlighted.

The paper considers a Gaussian response case. However, route map considerations are directly applicable to alternative response distributions via generalized GWR models (Fotheringham et al. 2002; Atkinson et al. 2003; Nakaya et al. 2005; Waller et al. 2007; Nakaya 2015; Dong et al. 2018; Comber et al. 2018a), or where the response is measured through a series of quantile-based distributions (Chen et al. 2012; 2020; Harris and Juggins 2011). They are similarly relevant to extended GWR models that include temporal considerations (Huang et al. 2010; Fotheringham et al. 2015; Du et al. 2018; Wu et al. 2019), contextualized GWR models dealing with hierarchical data (Harris et al. 2013), and GWR models that downscale outputs from area to point support (Murakami and Tsutsumi 2015; Jin et al. 2018).

## 2. Models and Data

### 2.1 Linear regression, standard, mixed and multiscale GWR

Although, various forms of fixed coefficient regression and varying coefficient GWR models will be referred to in this study, it is first useful to describe four models that are considered of *primary* importance to a GWR study. Here, the basic linear regression model can be defined as:

$$y_i = \beta_0 + \sum_{k=1}^{m} \beta_k x_{ik} + e_i \qquad (1)$$

where for observations indexed by $i=1...n$, $y_i$ is the response variable, $x_{ik}$ is the value of the $k^{th}$ predictor variable, $m$ is the number of predictor variables, $\beta_0$ is the intercept term, $\beta_k$ is the regression coefficient for the $k^{th}$ predictor variable and $e_i$ is the random error term that is independently normally distributed with zero mean and common variance $\sigma^2$. OLS is commonly used for model estimation in linear regression models.

Standard GWR is similar to linear regression but calibrates the regression model at point locations $(u, v)$ either from the sampled data or otherwise, using nearby sampled data falling within a moving window or kernel at the centre of each discrete location:

$$y_i = \beta_0(u_i, v_i) + \sum_{k=1}^{m} \beta_k(u_i, v_i) x_{ik} + e_i \qquad (2)$$

where $(u_i, v_i)$ is the spatial coordinate of the $i^{th}$ observation and $\beta_k(u_i, v_i)$ is a realization of the continuous function $\beta_k(u, v)$ at point $i$. As with the linear regression model, the set of $e_i$ obey an independent normal distribution with zero mean and common variance $\sigma^2$. In contrast to the *global* linear regression, GWR conducts *local* regression at any given location (the *geographical* part of GWR), using observations weighted by their distances to the location under consideration (the *weighted* part). Equations for calculating the local coefficient standard errors for GWR can be found in Fotheringham et al. (2002) and Harris et al. (2010a).

The weightings in GWR are determined by a kernel-based distance decay function and its bandwidth. Bandwidth can be a fixed distance or a fixed number of nearest data points (i.e. an adaptive radius depending on the local density of points). Automated routines exist to determine an optimal bandwidth by minimizing some measure of model fit such as the Akaike information criterion (AIC) and its corrected version (AICc) (Fotheringham et al. 2002, following Akaike 1973; Hurvich and Tsai 1989), Bayesian Information Criterion (BIC) (Nakaya 2001, following Schwarz 1978) or a leave-one-out cross validation score (CVS) (Brunsdon et al. 1996; 1998a, following Bowman 1984). As the bandwidth increases, the standard GWR estimator asymptotically converges to the OLS estimator of the *whole map* linear regression model.

In the standard form, a single bandwidth is used to calibrate GWR. This may be unrealistic because it implicitly assumes that each response-to-predictor relationship operates at the same spatial scale. Some relationships may operate at larger scales and others at smaller scales. A standard GWR will nullify these differences and find a 'best-on-average' scale of relationship non-stationarity. In this respect, mixed (or semiparametric) GWR (MX-GWR) (Brunsdon et al. 1999; Mei et al. 2004; 2006; 2016) can be implemented in which some relationships are assumed to be stationary whilst others are assumed non-stationary. However, MX-GWR only in part addresses the limitation of standard GWR, as the subset of locally varying relationships is still assumed to operate at the same spatial scale.

To fully address this, multiscale GWR (MS-GWR) (Yang 2014; Lu et 2017; 2018; Fotheringham et al. 2017; Leong and Yue, 2017; Yu et al. 2019; Oshan et al. 2019; Li et al. 2020) can be used, in which each relationship is specified using its own bandwidth, and the scale of relationship non-stationarity may vary for each response-predictor relationship. Unlike the linear regression and GWR, both MX-GWR and MS-GWR require an iterative back-fitting procedure for their estimation and as such can be computationally demanding (Lu et al. 2018; Li and Fotheringham 2020). Descriptions moving from GWR to MX-GWR and from GWR to MS-GWR, including calculations for coefficient standard errors and $t$-values, can be found in Mei et al. (2016) and Yu et al. (2019), respectively building on the initial work outlined in Yang et al. (2011).

In this study's implementations of GWR, MX-GWR and MS-GWR, a bi-square weighting kernel is used (e.g. see Gollini et al. 2015) where a single bandwidth $b$ is found for standard GWR and also for the pre-specified local or non-stationary relationships in MX-GWR, while $m + 1$ bandwidths are found for MS-GWR. All bandwidths are optimized by minimizing the AICc.

## 2.2 Case study data

The case study consists of a single soil dataset of 689 observations, spaced at approximately 100 m in a small watershed in the Loess Plateau, China (110.32821°E and 38.83433°N). The data are shown in Figure 1 and described in Wang et al. (2009) who undertook only a linear regression analysis but complemented with a geostatistical variographic analysis. The data are also described in Comber et al. (2018b) who used the same data to develop an extension to GWR. The data set includes soil total nitrogen (*STN*), taken as the response variable, and six predictor variables; soil organic carbon (*SOCgkg*), nitrate nitrogen (*NO3Ngkg*), ammonium (*NH4Ngkg*), and percentage clay (*ClayPC*), silt (*SiltPC*), sand (*SandPC*) content. In both Wang et al. (2009) and Comber et al. (2018b), the data were transformed, and this operation is retained here: *STN*, *SOCgkg*, *NO3Ngkg* and *NH4Ngkg* are transformed using natural logs and *ClayPC* is square root transformed. As with any regression analysis due consideration should be given to the nature of data relationships, the use of data transforms and associated model specification tasks prior to the main model fits – see Appendix A1.

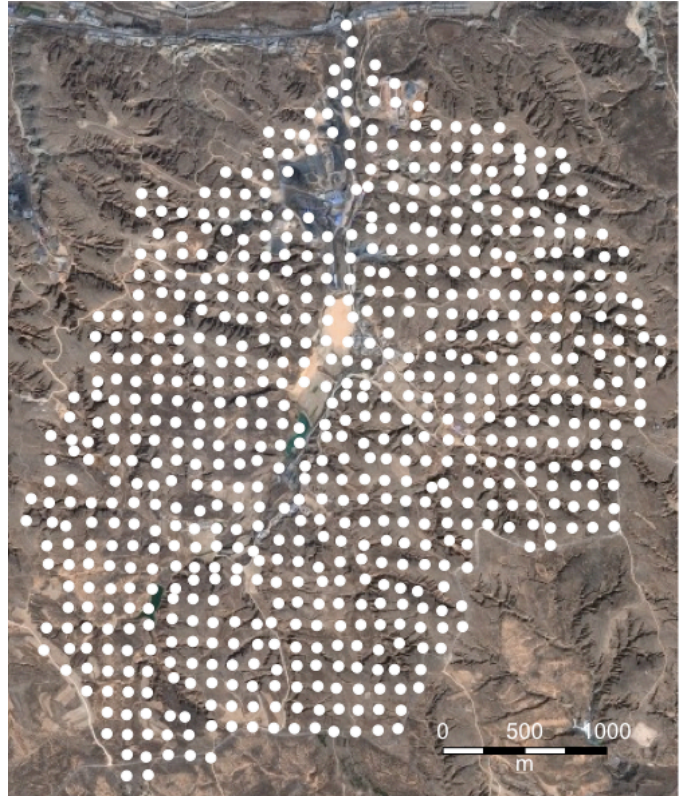

Figure 1. The case study data locations.

Each analysis in the GWR route map below predicts *STN* using different predictor variable subsets to illustrate specific points. At no point is the intention to conduct a nuanced regression analysis that attempts to fully characterise and interpret the soil processes. Rather the different data set scenarios are used only to illustrate the route map. The data set and the R code used to undertake the analyses are available from https://github.com/lexcomber/GWRroutemap.

## 2.3 Case study scenarios

Four data set scenarios were chosen to illustrate the route map decisions. These are given in Table 1, each with *STN* as the response but with differing predictors. The compositional nature of the clay / sand / silt data is catered for by omitting at least one from an analysis. Critically, the intention is to treat each scenario as a distinct and independent data set and not as a linked model specification exercise with respect to predictive variable selection. In this respect, 'Analysts' are assigned to each data set, where Analysts B-D are entirely unaware that more predictors of *STN* exist. To emphasise, this directly entails that model fit statistics such as AICc should not be compared across all study models (i.e. those of all four scenarios) but only compared for those models relating to each scenario, in turn.

| | *SOCgkg* | *ClayPC* | *SiltPC* | *SandPC* | *NO3Ngkg* | *NH4Ngkg* |
|---|---|---|---|---|---|---|
| **Analyst A** | yes | yes | yes | - | yes | yes |
| **Analyst B** | - | - | - | yes | yes | - |
| **Analyst C** | yes | - | - | - | - | yes |
| **Analyst D** | yes | - | - | yes | yes | - |

Table 1. Data set scenarios in terms of four different 'Analysts'.

## 3. Primary model decisions

The fundamental consideration for undertaking a GWR analysis is that it should be justified in terms of the aims of the analysis and the characteristics of the data. If spatial effects are evident in the data (see Appendix A1 for data considerations and exploratory mapping of variables) then a GWR can be considered but this requires demonstrating that alternate models, specifically ones with fixed coefficients, are not suitable. To achieve this, the following steps for any GWR analysis are recommended:

1) A basic linear regression should be undertaken and the results investigated.
2) A MS-GWR should be calibrated and the estimated bandwidths interrogated.
3) Based on findings (1) and (2), one from a standard GWR, MX-GWR and MS-GWR should be considered for further analysis *provided* an SVC model is considered suitable in the first place.

The linear regression model assumes fixed data relationships and provides the baseline against which all forms of GWR can be compared. The MS-GWR model, estimates the bandwidths for each response-predictor relationship. Evaluating these directly quantifies any spatially varying relationships and at what spatial scale they each operate at. This in turn informs on whether to pursue a GWR analysis and if so, which of three different GWR forms to follow. That is, given the MS-GWR results, can a simpler model in a linear regression, standard GWR or MX-GWR provide a viable and pragmatic alternative? Or is MS-GWR the only viable option?

This approach to *primary* model choice is recommended first because the theory for the standard linear regression is extremely well developed, whilst theoretical developments reduce exponentially moving from standard GWR, to MX-GWR, and finally to MS-GWR, where MS-GWR is relatively recent with some theoretical consideration still unresolved (Lu et al. 2019). Second, other critical considerations of model complexity, sample size, sample configuration and sample variation play key and intertwined roles, which cannot be entirely resolved through a comparison of model parsimony-fit statistic such as AIC / AICc. Thus, choosing a simpler regression over the relatively complex MS-GWR is advocated but where this decision is informed by following the proposed route map.

Both the linear regression and MS-GWR analysis should also investigate for the presence of spatially autocorrelated model residuals. Thus, further to the four model choices (of linear regression, GWR, MX-GWR and MS-GWR), a fifth model is considered where an alternative fixed coefficient regression is fitted but with a spatially autocorrelated error term (i.e. a spatially autocorrelated model, SAM). For this study, the spatially autocorrelated error term is modelled by the parameterization of its covariance using an exponential function decaying with respect to the Euclidean distance separating sample sites. The restricted maximum likelihood (REML) method (e.g. Lark et al. 2006) is used for the estimation. The SAM will warrant consideration depending on the nature of spatially autocorrelated residuals from the linear regression fit and also if the MS-GWR fit indicates that only the intercept is found to be spatially varying (Nakaya et al. 2005; Harris et al. 2010b; Harris 2019). Again, the theory for the SAM and related models is well developed (e.g. Schabenberger and Gotway 2004; Waller and Gotway 2004).

The kernel bandwidth identification is *the* critical consideration in GWR as it determines how many data points are included in the data subset passed to each local regression and how these data points are spatially weighted. Bandwidths dictate the degree of smoothing or variation in the local regression coefficient estimates, and the study's interpretations and inferences for process heterogeneity thereafter.

Determining the scale at which data relationships operate is not a straightforward task. In this study, bandwidths are found objectively via AICc, but this should not discount user-specified bandwidths when there exists some strong prior belief, theoretical justification or expert knowledge for their use. Similar discussions can be found in related kernel weighting paradigms, such as kernel density estimation, where automated bandwidth approaches are not necessarily viewed as a panacea for bandwidth selection (Silverman, 1986). There are strong benefits in conducting an extensive bandwidth investigation, as final model outputs are more assured.

For the *primary* analyses, only rudimentary assessments of statistical (relationship) significance are undertaken using coefficient standard errors, *t*-values and *p*-values from standard GWR, MX-GWR and MS-GWR models. Caveats on their use with all forms of GWR are discussed in Appendix A3.

### 3.1 Step 1: Basic linear regression and autocorrelated residuals

The first step is to undertake a global linear regression. The aim for the regression analysis is to try to understand how the predictors relate to the response variable, specifically: (a) which relationships are statistically *significant*, (b) evidence for specifying an *autocorrelated error* term, and (c) *the fit* of the linear regression itself. Table 2 summarises the linear regression coefficient estimates and their significance from zero, for all four Analysts. The linear models from Analysts A and C provide a mixture of significant and insignificant predictors of *STN*, while all predictors are significant for the linear models from Analysts B and D.

| | **Analyst A** | | **Analyst B** | | **Analyst C** | | **Analyst D** | |
|---|---|---|---|---|---|---|---|---|
| | **Estimate** | ***p*-value** | **Estimate** | ***p*-value** | **Estimate** | ***p*-value** | **Estimate** | ***p*-value** |
| *Intercept* | -2.220 | 0.000 | -0.723 | 0.000 | -2.130 | 0.000 | -1.437 | 0.000 |
| *SOCgkg* | 0.690 | 0.000 | - | - | 0.918 | 0.000 | 0.683 | 0.000 |
| *ClayPC* | -0.011 | 0.843 | - | - | - | - | - | - |
| *SiltPC* | 0.015 | 0.000 | - | - | - | - | - | - |
| *SandPC* | - | - | -0.021 | 0.000 | - | - | -0.012 | 0.000 |
| *NO3Ngkg* | 0.126 | 0.000 | 0.355 | 0.000 | - | - | 0.112 | 0.000 |
| *NH4Ngkg* | -0.146 | 0.047 | - | - | -0.011 | 0.884 | - | - |

Table 2. Linear regression coefficient estimates and their significance (*p-value*).

To assess spatial autocorrelation of the linear regression residuals, a spatial weight matrix was defined and unbiased estimates of Moran's *I* and their significance were determined (Table 3), under the expectation of random and independent residual distributions. Moran's *I* for all four models are significant, where the spatial structure in the linear regression residuals varies from relatively weak to relatively strong (data for Analyst A has the weakest structure, while data for Analyst C has the strongest), as reported by the magnitude of the estimates. In this case, all four data set scenarios indicate that a fixed coefficient regression with a spatially autocorrelated error term could be suitable (i.e. a SAM via a REML estimation). This is not surprising given the data are spatial.

Table 3 also summaries the error statistics for the four scenarios with AICc and $R^2$ values. AICc is an *in-sample* statistic reflecting model parsimony, while $R^2$ is also *in-sample*, reflecting model prediction accuracy. Ideally, an *out-of-sample* statistic should also be reported, such as the CVS or the PRESS statistic (Allen 1974), as this addresses a certain bias found with *in-sample* statistics. However, no currently coded MS-GWR model could provide such an *out-of-sample* measure.

| | Moran's *I* | *p*-value | AICc | $R^2$ |
|---|---|---|---|---|
| **Analyst A** | 0.142 | 0.000 | 1124.0 | 0.609 |
| **Analyst B** | 0.174 | 0.000 | 1377.4 | 0.430 |
| **Analyst C** | 0.219 | 0.000 | 1223.1 | 0.545 |
| **Analyst D** | 0.144 | 0.000 | 1131.0 | 0.603 |

Table 3. Residual autocorrelation measures using Moran's *I* and fit statistics (AICc and $R^2$) from the four linear regression fits.

Thus, for all four scenarios, there is no indication, as of yet, that a GWR analysis may be appropriate, although the existence of autocorrelated residuals from a linear regression fit commonly *suggests* that a GWR analysis may be useful, even though such outcomes do not indicate the presence of spatially varying relationships between the response and the predictor variables. This observation is critically important for understanding spatial regression modelling in general and is routinely confused in GWR studies. Useful discussions on this misconception, together with issues of identifying spatial autocorrelation effects from spatial heterogeneity in terms of regression relationships can be found in Harris (2019) and references therein.

In general, but not a rule, measures of strong model fit (e.g. $R^2 > 0.8$), coupled with weak and insignificant levels of spatial autocorrelation in the residuals, suggest that a linear regression would be appropriate. This might be a fully specified model that included measures of all likely predictors or factors driving the soil response variable, some of which are inherently spatial (e.g. topography, soil class, etc.). If the fit is poor and exhibits significant levels of residual spatial autocorrelation, a GWR analysis is still an option, as is a SAM.

In summary, this first step has fitted a linear regression model to identify which relationships are globally significant and whether spatial autocorrelation effects may potentially exert an important influence on these findings.

## 3.2 Step 2: Multiscale GWR (MS-GWR) and bandwidth estimation

The second (and concurrent) step of the *primary* route map is to undertake an MS-GWR analysis. This informs on the different scales of relationships in the data, where some may be local and others global. The MS-GWR bandwidths explicitly describe the degree of spatial heterogeneity associated with each variable's relationship to the response. For the MS-GWR analysis at this stage, only the following need investigation: (i) the *estimated bandwidths* (ii) evidence for *residual autocorrelation*, and (iii) *the fit* of MS-GWR itself.

The estimated MS-GWR fixed distance bandwidths are shown in Table 4, with adaptive distance bandwidths illustrated for the MS-GWR model of Analyst A only. In this study, the maximum number of data points that can be included under an adaptive bandwidth is 689 (the total number of observations in the data) and the maximum fixed bandwidth is 3742 m (the maximum distance between any pair of data points). The bandwidths in Table 4 should be interpreted in light of these values. For Analyst A, the order of bandwidth size is consistent between fixed and adaptive forms, and this was broadly the case for the other three data set scenarios. This similarity is re-assuring but to a certain extent reflects that the study data were sampled on a loosely regular grid. Studies with data clearly on an irregular sample configuration may need to experiment more in this respect (see Appendix A1).

|  | Intercept | SOCgkg | ClayPC | SiltPC | SandPC | NO3Ngkg | NH4Ngkg |
|---|---|---|---|---|---|---|---|
| **Analyst A** | 555.9 | 2483.9 | 3741.7 | 1080.8 | - | 382.5 | 3741.7 |
| **Analyst A\*** | 57 | 631 | 685 | 306 | - | 55 | 685 |
| **Analyst B** | 445.8 | - | - | - | 1232.9 | 731.9 | - |
| **Analyst C** | 424.9 | 3741.4 | - | - | - | - | 3741.8 |
| **Analyst D** | 573.6 | 2214.6 | - | - | 1066.5 | 378.4 | - |

Table 4. The fixed bandwidths in metres (max = 3742 m) for different models arising from an MS-GWR. For Analyst A, * indicates an adaptive bandwidth (max = 689).

On viewing the fixed bandwidth results only, clear patterns emerge relating to each predictor variable and the scale of its spatially varying relationship to the response, *STN*. For Analyst A, the MS-GWR bandwidths for *ClayPC* and *NH4Ngkg* both strongly tend towards the maximum, global bandwidth of 3742 m, while *SOCgkg* and *SiltPC* have bandwidths of about two-thirds and one-third of the global one, respectively. The bandwidths for the *intercept* and *NO3Ngkg* for Analyst A are both strongly local. For Analyst B, the bandwidths for the *intercept, SandPC and NO3Ngkg* are all local. For Analyst C, the bandwidths for *SOCgkg* and *NH4Ngkg* are essentially global, while the *intercept* is local. For Analyst D, none of the bandwidths appear global, where those for the *intercept, SOCgkg, SandPC and NO3Ngkg* vary locally but appear quite different in magnitude.

To assess residual spatial autocorrelation for the MS-GWR fits for each Analyst, estimates of Moran's *I* and their significance are given in Table 5, along with MS-GWR fit statistics. For each Analyst only, the results need to be directly compared to the corresponding results given in Table 3 (for the linear regression model). Clearly in all data set scenarios, residual autocorrelation is now negligible, while model fit improves over that found for the corresponding linear regression. Note that the Moran's *I* analysis for MS-GWR does not account for first- to second-order identification bias (see Armstrong 1984), unlike the bias accounted for in the corresponding Moran's *I* analysis for the linear regressions in Table 3.

|  | Moran's *I* | *p*-value | AICc | *R*$^2$ |
|---|---|---|---|---|
| **Analyst A** | -0.007 | 0.604 | 1050.4 | 0.713 |
| **Analyst B** | -0.013 | 0.700 | 1264.4 | 0.580 |
| **Analyst C** | 0.005 | 0.381 | 1106.8 | 0.662 |
| **Analyst D** | -0.009 | 0.636 | 1057.4 | 0.708 |

Table 5. MS-GWR residual autocorrelation measures using Moran's *I* and error statistics.

## 3.3 Step 3: Choice of *primary* model

The results of the initial linear regression and MS-GWR analyses guide *primary* model choice. First, from the linear regression analysis in Step 1, it appears that a fixed coefficient model should only be considered if calibrated with an autocorrelated error term, for all four Analysts (i.e. all can consider SAM fits). Second, from the MS-GWR analysis in Step 2, some form of GWR is similarly worth considering, for all Analysts, as residual autocorrelation essentially disappears with a MS-GWR fit, while at least one predictor bandwidth, including that for the intercept, is clearly local. Step 2 MS-GWR models also consistently improve fit over their corresponding linear models of Step 1. For deciding on the *primary* model, the following sub-sections provide guidance to how this should be undertaken considering each of five regression possibilities (Linear regression,

SAM, standard GWR, MX-GWR and MS-GWR) and the four data set scenarios. Critical to Step 3 are the presentation and interpretation of the estimated coefficients and associated uncertainties from competing models.

### Investigating linear regression and SAM for Analyst C

A linear regression should be considered as a potential final model when all bandwidths from MS-GWR are large (i.e. tend towards the global situation), including the intercept. As a rule of thumb, this is when they are broadly greater than 80% of the maximum distance between data points (or 80% of the data points in the adaptive bandwidth case). In this respect, none of the Analysts have a data set that clearly suggests a linear regression fit to be appropriate. However, from above, it is stated that all analysts could consider a SAM fit (as all indicated autocorrelated residuals from their linear regression models), and in this respect, a SAM can be further endorsed if all predictor variable bandwidths from MS-GWR tend to the global, but the *intercept* is local. This is clearly the case for Analyst C's data set (from Table 4).

Thus, in this instance, the *primary* route map has guided Analyst C to a SAM. It is prudent to compare SAM outputs to the linear regression model outputs because only the intercept term is locally varying from the MS-GWR. The coefficient summaries in Table 6 indicate only marginal gains in process interpretation with the SAM fit. However, the AICc improves with the SAM (1148.4 compared to 1223.1 for the linear regression). Thus, in this instance, there is only marginal inferential value to the inclusion of second-order spatial effects via a SAM, as reflected by the broadly similar estimates and statistical inferences of regression coefficients to the non-spatial linear regression. Analyst C could also have considered an MX-GWR with only the *intercept* locally varying, but as a rule, spatial effects via a SAM should always be preferred due to its stronger inferential properties. Thus, in summary, Analyst C should proceed with a fixed coefficient regression, where a linear regression suffices.

| | Linear regression | | SAM | |
|---|---|---|---|---|
| | **Estimate** | ***p*-value** | **Estimate** | ***p*-value** |
| *Intercept* | -2.130 | 0.000 | -1.817 | 0.000 |
| *SOCgkg* | 0.918 | 0.000 | 0.816 | 0.000 |
| *NH4Ngkg* | -0.011 | 0.884 | -0.086 | 0.284 |

Table 6. Coefficient estimates and their significance arising from linear regression and SAM fits for Analyst C.

### Investigating MX-GWR and MS-GWR for Analyst A

An MX-GWR can be experimented with when the MS-GWR analysis suggests two distinct sets of bandwidths, with one set tending to the global and the other set tending to a similar local scale. This scenario appears likely for Analyst A (from Table 4), where the MS-GWR bandwidths for *SOCgkg, ClayPC* and *NH4Ngkg* can be viewed as global, while those for the *intercept*, *SiltPC* and *NO3Ngkg* can be viewed as local.

As example, an MX-GWR is fitted to Analyst A's data set. Figure 2 shows the spatial distribution of the local coefficient estimates with those significantly different to zero highlighted (i.e. *p*-values < 0.05). The local coefficients portray the geographically varying relationships between the *intercept*, *SiltPC* and *NO3Ngkg* to *STN*, where the *NO3Ngkg* relationship can change in sign. The global coefficient estimates for *SOCgkg ClayPC* and *NH4Ngkg* from the MX-GWR fit were 0.677, -0.016 and -0.193, respectively (where that estimated for *ClayPC* was the only one not significantly different to zero). To fit the MX-GWR model, a single local bandwidth needs to be determined, and in this instance, it was user-specified to be 700 m. The MS-GWR coefficient estimates should also be mapped for comparison and are given in Figure 3. Further, the AICc fit of the MX-GWR model was estimated to be poorer at 1065.9 to that found with MS-GWR at 1050.4.

Thus, to fully interpret the nature of the relationships in Analyst's A data set, coefficient summaries found for linear regression (Table 2), MX-GWR (Figure 2) and MS-GWR (Figure 3) need to be jointly considered. On balance, *STN*'s relationships with *SOCgkg*, *ClayPC* and *NH4Ngkg* are clearly global and constant across space, where *STN*'s relationships with *ClayPC* and *NH4Ngkg* are not viewed as significant, noting that the *NH4Ngkg* relationship to *STN* is borderline significant / insignificant in all fits (linear regression, MX-GWR and MS-GWR). Conversely, *STN*'s relationship with the *intercept*, *SiltPC* and *NO3Ngkg* are local, where the local behaviour varies little between the MX-GWR and MS-GWR forms. Only for *NO3Ngkg* do differences occur, where more distinct and significant areas of negative coefficient estimates were generated with MS-GWR, but not seen in MX-GWR. If the differences were more pronounced, then Analyst A should consider re-specifying the MS-GWR model with bandwidths for *SOCgkg*, *ClayPC,* and *NH4Ngkg* pre-set (or fixed) as global, while those for the *intercept*, *SiltPC* and *NO3Ngkg* re-estimated so that each relationship varies at its own local scale. However, given the similarity in the coefficient distributions, Analyst A could justifiably and pragmatically proceed with a MX-GWR fit, even with its worse AICc.

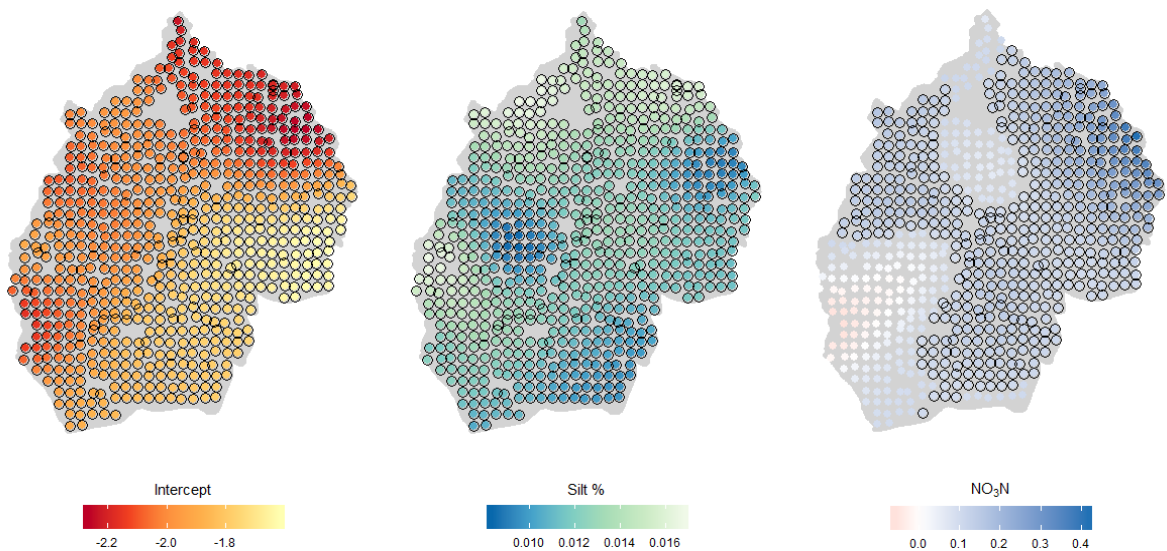

Figure 2. The spatial variation of the local coefficient estimates given with *p*-values < 0.05 highlighted from the MX-GWR analysis of Analyst A.

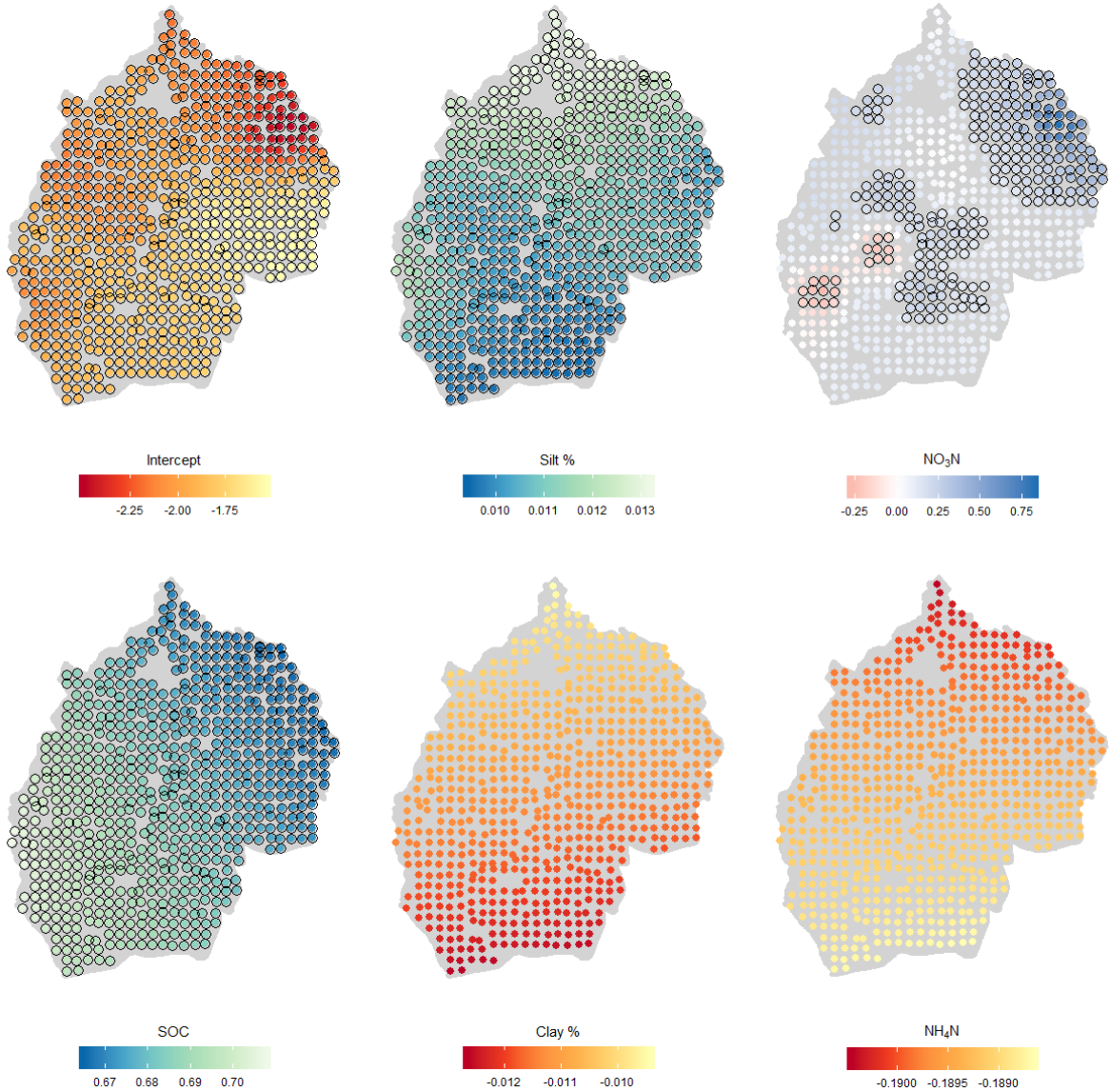

Figure 3. The spatial variation of the local coefficient estimates given with *p*-values < 0.05 highlighted from the MS-GWR analysis of Analyst A.

### *Investigating MS-GWR only for Analyst D*

The MS-GWR fit should be retained when its bandwidths clearly suggest each data relationship is operating at its own unique spatial scale. Here the data set for Analyst D provides such an instance (see Table 4), where Figure 4 maps the distribution of the local coefficient estimates. Here, only the relationship for *NO3Ngkg* with *STN* changes in sign; and is the only relationship that geographically varies between significant and insignificant.

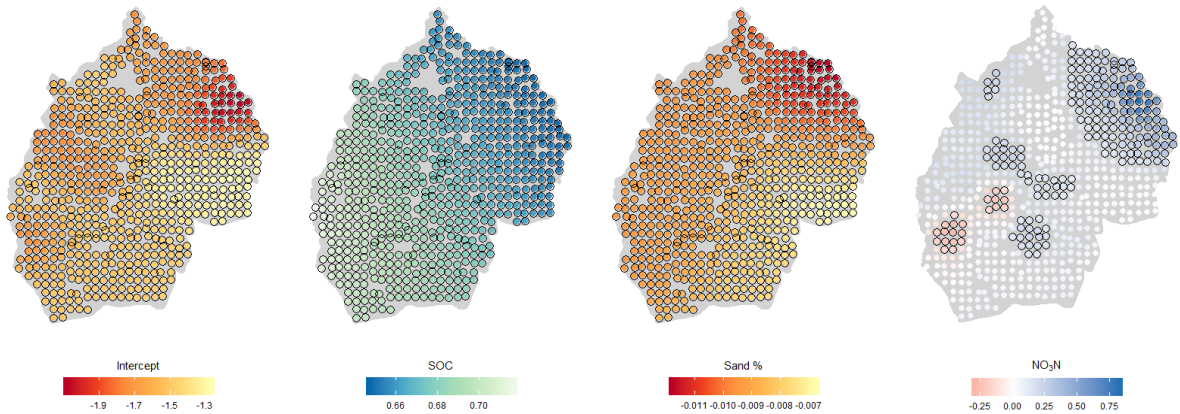

Figure 4. The spatial variation of the local coefficient estimates given with *p*-values < 0.05 highlighted from the MS-GWR analysis of Analyst D.

### *Investigating standard GWR and MS-GWR for Analyst B*

A standard GWR is generally not an adequate model. It can be chosen over an MS-GWR only on the rare occasions when the intercept and all predictors have broadly similar MS-GWR estimated bandwidths, potentially as that found for Analyst B (Table 4). This scenario predicts *STN* using just *SandPC* and *NO3Ngkg*, for which a single local bandwidth appears reasonable. In this instance, the single bandwidth can be optimally determined through a standard GWR calibration, where it was found via AICc to be 597.5 m.

Where possible, the bandwidth function in standard GWR should be investigated, and can be considered analogous to an investigation of the variogram in Geostatistics, where both investigations aim to identify spatial structure in some way (e.g. Cressie 1989). This ensures that the bandwidth optimisation has not settled on a local minimum and allows the degree to which the identified bandwidth is optimal to be confirmed. Figure 5 shows the bandwidth function for an AICc minimisation, which is well-behaved with a clear minimum. Observe that if the bandwidth function was very shallow and plateaued, then a linear regression would likely suffice. Also, small bandwidths (say, < 2% when using an adaptive bandwidth) are indicative of over-fitting, and that a standard GWR is suggesting geographical patterns when none exists. In this case, the GWR analysis should cease. The problem of over-fitting in standard GWR is well known (e.g. Jetz et al. 2005; Páez et al. 2011), but GWR also has the capacity to under-fit (Harris 2019).

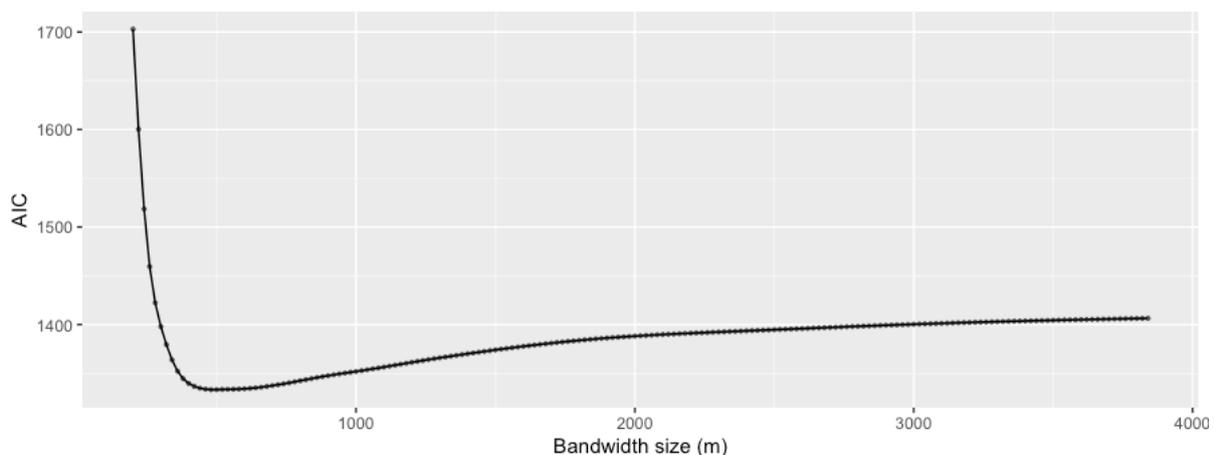

Figure 5. The bandwidth function for standard GWR.

Figure 6 maps the distribution of the local coefficient estimates from standard GWR. Here, the relationships for the *intercept* and *SandPC* with *STN* can change in sign. Again, the MS-GWR coefficient estimates are mapped for comparison (Figure 7), indicating clear spatial differences between standard GWR and MS-GWR coefficients. In general, MS-GWR indicates smaller ranges of coefficient variation, but where the regression relationships are consistently significant across space. Thus, given these differences and that the AICc for standard GWR is poorer at 1272.3 to that found with MS-GWR at 1264.4, it is considered prudent to retain the MS-GWR model rather than simplifying the analysis with standard GWR.

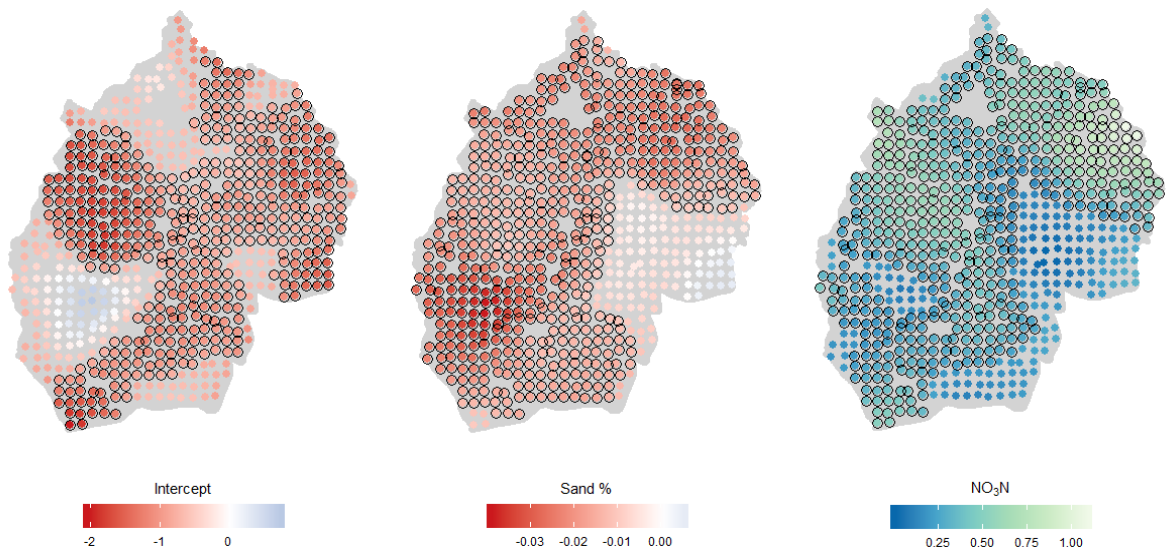

Figure 6. The spatial variation of the local coefficient estimates given with *p*-values < 0.05 highlighted from the standard GWR analysis of Analyst B.

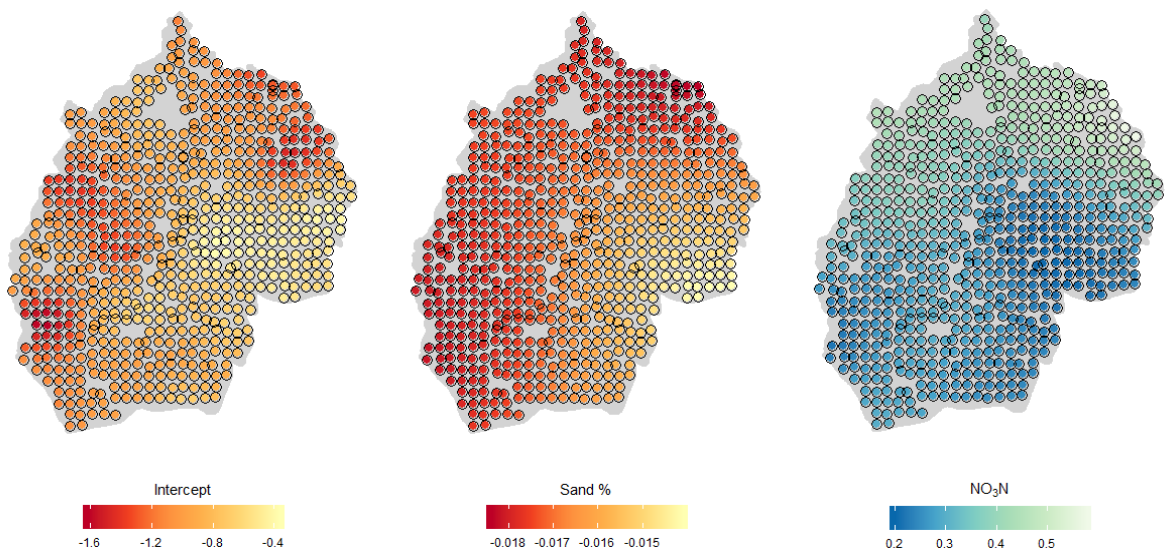

Figure 7. The spatial variation of the local coefficient estimates given with *p*-values < 0.05 highlighted from the MS-GWR analysis of Analyst B.

*Summary in terms of AICc*

Table 7 summarises the AICc results for each Analyst, where for all data set scenarios, the MS-GWR model provides the most parsimonious fit in terms of AICc. The chosen *primary* model is always an improvement in fit over the linear regression model but does not necessarily provide an improvement in fit over the corresponding MS-GWR model in terms of AICc. This is because the interpretations of relationship non-stationarity (via the coefficient maps, above) can sometimes remain broadly unaltered when a poorer fitting but relatively simple model (e.g. MX-GWR) is specified rather than the relatively complex MS-GWR model.

|  | **Linear regression** | **MS-GWR** | **Primary model 'chosen'** |
|---|---|---|---|
| **Analyst A** | 1124.0 | 1050.4 | 1065.9 (MX-GWR) |
| **Analyst B** | 1377.4 | 1264.4 | 1264.4 (MS-GWR) |
| **Analyst C** | 1223.1 | 1106.8 | 1223.1 (Linear regression) |
| **Analyst D** | 1131.0 | 1057.4 | 1057.4 (MS-GWR) |

Table 7. AICc values arising from the *primary* model analyses.

## 4. Discussion of *secondary* model decisions

Having arrived at *GWR Basecamp* through a *primary* analysis and where one from a standard GWR, MX-GWR or MS-GWR form is considered suited to the observed spatially varying relationships, the second stage of the GWR route map is the consideration of *secondary* GWR model issues. As stated in the introduction, strategies for *secondary* model decisions (*scaling the summit*) are only described and not implemented (through the case study data sets).

In order of importance, the following issues should be investigated: (a) *predictor collinearity*, (b) the *influence of outliers*, and (c) evidence of a *dependent error term*. These should be examined at both global (as indicated in Appendix A1) and local contexts, but here the focus needs to be placed locally with the associated GWR form. As with any fixed coefficient, global regression, these issues can be similarly detrimental to a reliable GWR analysis, giving rise to say, spurious local changes in the sign of the coefficient estimates between positive and negative and local changes in significance. They can also compromise bandwidth estimation, where GWR fits of a *secondary* analysis will often give rise to different (optimised) bandwidths or a change in the behaviour of the bandwidth function to that found with the *primary* analysis, and thus, potentially changing the chosen GWR form (e.g. see respectively, Gollini et al. 2015; Harris et al. 2010a; Cho et al. 2010).

Crucial to the *secondary* analysis is to test for the presence of each issue and if found, to determine the impact of the issue and the potential for misinterpreting the GWR model outputs from the *primary* analysis. Thus, comparisons of GWR diagnostic statistics and coefficient outputs need to be made, although unfortunately this can become problematic when the given *secondary* analysis GWR form does not exist (either in concept or in code). For example, a robust (outlier-resistant) MS-GWR has not, as yet, been developed. A further complication arises, when multiple issues are observed (e.g. local collinearity and local outliers), where say, a robust ridge GWR model might be suitable but is not available.

Given these difficulties, some simple global data processing operation may sometimes resolve multiple issues, say through a data transform to negate the use of a robust GWR model and / or to address any heteroskedastic error effects (see Appendix A1). Furthermore, the choice of the *primary* GWR model may indirectly negate a *secondary* issue. For example, both MX-GWR and MS-GWR has been empirically shown to indirectly account for local collinearity observed in standard GWR (Geniaux and Martinetti 2018; Harris 2019).

*Collinearity*

For any global regression, collinearity occurs when pairs of predictors have a strong linear relationship between each other, either positive or negative. Broadly, collinearity may be a problem when correlation coefficients for a predictor pair are > 0.8 or < -0.8 as these can affect model reliability and precision. Diagnostics such as matrix condition numbers (CNs), predictor variance inflation factors (VIFs) and variance decomposition factors (VDPs) can be found where rules of thumb can be applied (CNs > 30, VIFs > 10 and VDPs > 0.5) to indicate worrying levels of collinearity (Belsey et al. 1980). Often a simple remedy is to remove one or more predictors. The difficultly is in deciding which predictor(s) to remove, especially when all are considered important to describing the study process. Here, a penalized regression can provide a sophisticated solution, that by design includes a model specification capability (Zou and Hastie 2005; Friedman et al. 2010; Dormann et al. 2013).

Collinearity may also be present in some local predictor data subsets of GWR even when not observed globally (Wheeler and Tiefelsdorf 2005). Compositional, categorical and ordinal predictors can be particularly problematic, often resulting in exact local collinearity making bandwidth optimisation impossible. Geographically weighted collinearity diagnostics (CNs, VIFs and VDPs) are available for GWR (Wheeler 2007; 2013; Lu et al. 2014) and provided any observed collinearity is considered a concern (e.g. see the presentations of Páez et al. 2011; Fotheringham and Oshan 2016; Harris 2019), a standard GWR can be replaced with a penalized GWR form (Wheeler 2007; 2009; Brunsdon et al. 2012; Barcena et al. 2014; Gollini et al. 2015; Wang and Li 2017; Li and Lam 2018). Mapping geographically weighted correlation coefficients (Fotheringham et al. 2002; Harris et al. 2014) between predictor variable pairs can also be useful to identify areas of local collinearity.

*Outliers*

For outliers, it is first useful to examine the linear regression and MS-GWR residuals of the *primary* analysis for evidence of outliers that may influence the validity of their fits. This should be done spatially (with maps of standardized residuals, say), to determine where any GWR analysis may be compromised. Again, robust theory in the global case (e.g. Huber 1981; Marazzi 1993) has been transferred to the local case with robust extensions to standard GWR only (Fotheringham et al. 2002; Farber and Paez 2007; Harris et al. 2010a; Zhang and Mei 2011; Chen et al. 2012; Leyk et al. 2012; Lu et al. 2014). These handle influential outliers arising globally, but also locally in each individual regression, which may go undetected in any global assessment (i.e. via the standardized residual maps, above).

*Dependence in the error data*

As with linear regression estimated by OLS, most forms of GWR assume that the errors, $e_i$ are independently normally distributed with zero mean and common variance $\sigma^2$. To examine for a non-constant error variance (in a non-spatial, global manner), the regression's fitted values can be plotted against its residuals. A funnel shape indicates that a heteroskedastic regression should be considered, such as through some consistent estimator (see Davidson and MacKinnon 1993) or a weighted least squares (WLS) estimator. A direct extension to standard GWR is given in Fotheringham et al. (2002), where the error variance varies geographically. This heteroskedastic GWR form has also been developed to detect local outliers (Harris et al. 2010a) and to provide localised prediction variances (Harris et al. 2011, also see Appendix A4). Páez et al. (2002a; b) also provide a spatially heteroskedastic form of GWR but within a parametric framework, while Shen et al. (2011) extend the locally linear GWR model of Wang et al. (2008) to a heteroskedastic form.

Although it is common for any GWR fit to reduce error spatial autocorrelation over that found with a linear regression fit (as demonstrated in Section 3), it is likely that error autocorrelation will also occur for each local regression in a GWR. GWR models that account for local autocorrelation effects have been proposed including an extension to standard GWR (Brunsdon et al. 1998b; Cho et al. 2011) and an extension to MX-GWR (Geniaux and Martinetti 2018) through autoregressive GWR model forms.

## 5. Concluding remarks

Geographically Weighted Regression provides a framework to investigate spatial relationships in data, their heterogeneities and varying scales of interaction. Its use in analyses of environmental and socio-economic data continues to grow and is easily undertaken in a number of software implementations. However, an increasing number of GWR analyses reported in the literature are not appropriate to the study objectives or correctly formulated: in some cases GWR should not have been applied to the problem, in others the GWR model is incorrectly parameterised or the incorrect form of GWR is applied. Such situations may result in partial, incomplete or unreliable analyses and inference.

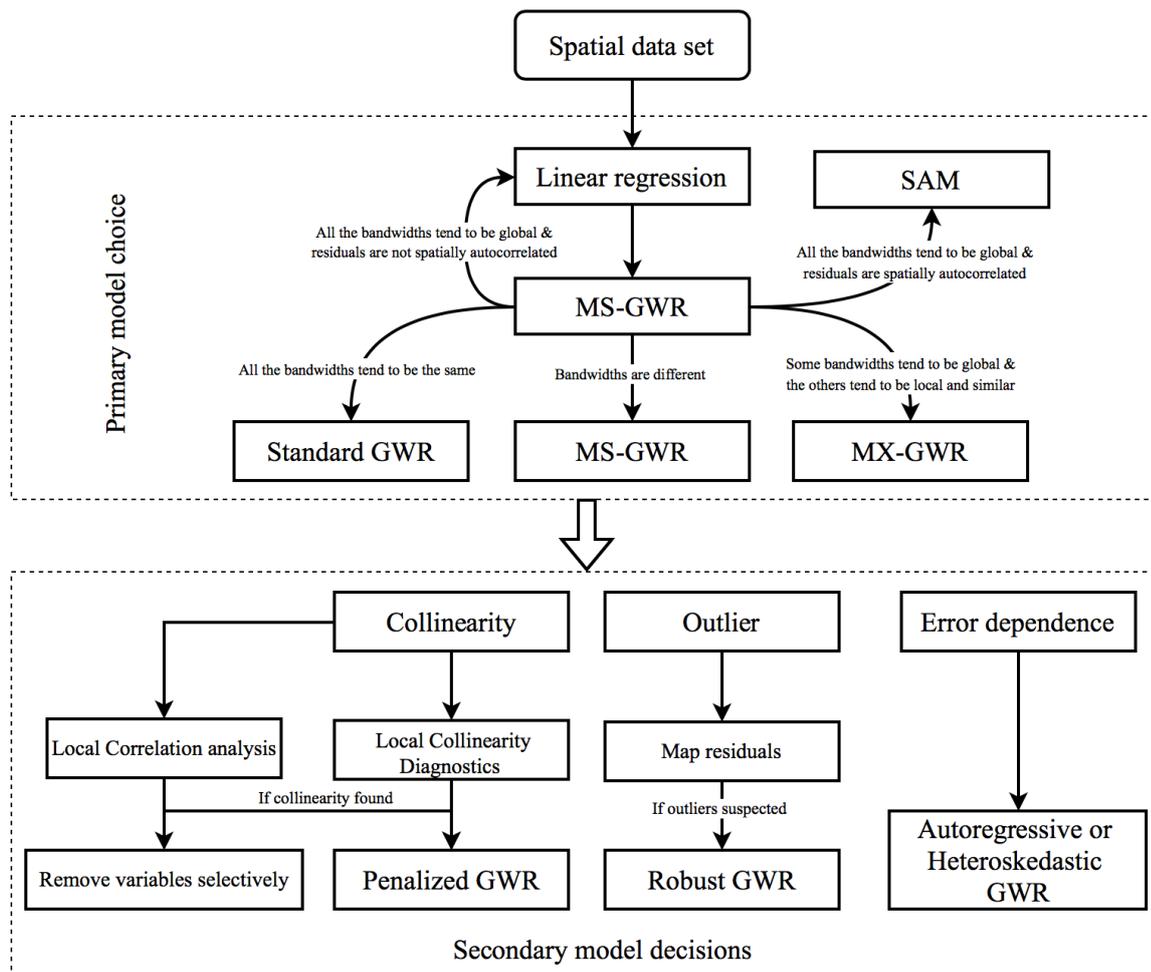

Figure 8 Flowchart of the GWR route map.

This paper describes a GWR route map of *primary* and *secondary* considerations to ensure the GWR analysis is justified in terms of the aims of the analysis and the characteristics of the data, over alternate models, with fixed regression coefficients. As summarized in Figure 8, the route map has the following *primary* steps:

1)      A linear regression analysis should always be undertaken and the results investigated.
2)      A MS-GWR (multi-scale GWR) should always be calibrated and the estimated bandwidths interrogated.
3)      Following the investigations of steps (1) and (2), the analysis should proceed with a standard GWR, or a core variant in MX-GWR (mixed GWR) or MS-GWR, only if a spatially varying coefficient model is considered appropriate. Otherwise a linear regression or a SAM (spatially autocorrelated model) should be chosen.

The linear regression (step 1) provides global insight into how the predictors relate to the response, which relationships are significant and measures of model fit. This step includes evidence of spatial autocorrelation in the residuals, for example through a Moran's *I* analysis.

The MS-GWR (step 2) provides information through the MS-GWR bandwidths about the different scales of relationships in the data, where some may be local and others global. The MS-GWR bandwidths describe the degree of spatial heterogeneity associated with each variable's relationship to the response. Insignificant Moran's *I* estimates of the spatial autocorrelation of the MS-GWR residuals provide evidence that accounting for relationship spatial heterogeneity using MS-GWR is capturing most of the structural variation in the data.

Investigations of the linear regression and MS-GWR results (step 3) guide the choice of the final *primary* model (i.e. a linear regression or SAM, standard GWR, MX-GWR or MS-GWR). A linear regression model should be retained when all bandwidths from MS-GWR tend towards the global situation, including the intercept (i.e. are greater than ~80% of the maximum distance between data points or 80% of the data points in the adaptive bandwidth case), and where spatial autocorrelation in the residuals is either absent or if present, does not significantly effect process interpretation (as the case for Analyst C, above). In many fixed coefficient cases however, instances of significant residual spatial autocorrelation are more likely to result in choosing a SAM over the non-spatial linear regression.

If spatial autocorrelation in the residuals is present and MS-GWR bandwidths are not all large, then a GWR variant can be considered:
-       A standard GWR should be considered in the rare situation when all of the MS-GWR bandwidths tends to the same value;
-       A MX-GWR should be considered when the MS-GWR bandwidths indicate two distinct sets of bandwidths, with one set tending to the global and with the other set tending to a similar local scale;
-       A MS-GWR should be considered when the all of bandwidths vary, suggesting that each data relationship operates at different spatial scales.

It is important to stress that the final model choice should not be guided by simply selecting the model with lowest AICc value, especially as the aim of any GWR analysis is to explore relationship spatial heterogeneity and spatial variations in process. Rather, interrogation of the coefficient estimates and their uncertainty arising from the different models is paramount. This point is somewhat philosophical in that the underlying assumption in model selection is the existence of the 'best model', as measured by AICc, say. All depends on the aims of conducting a GWR analysis in the first place, where for this study, relationship inference is the clear aim. However, if the study aim was for spatial prediction and associated inference with GWR, a very different route map would have been presented (see Appendix A4), together with associated bias-variance trade-offs.

This paper's GWR route map first provides a path through a number of *primary* issues and acts as a gateway to informed applications of GWR. The *primary* issues were demonstrated empirically through a soils data case study. As the next step of the route map, a number of *secondary* issues should be investigated once a GWR analysis has been decided upon. These *secondary* investigations focus at the local scale, including local predictor collinearity, the local influence of outliers, and local dependent error terms. *Secondary* considerations may interact with each other and with *primary* considerations and their investigation will further guide the decision to undertake a GWR analysis or not and if so, ensure an informed choice of which GWR form to use. Finally, further guidelines have been given in the Appendix, many of which can be of equal importance to those given in the main text.

**Acknowledgements**
This research was supported by the Natural Environment Research Council Newton Fund Grant (NE/N007433/1 and NE/S009124/1), the Biotechnology and Biological Sciences Research Council Grants (BBS/E/C/000J0100, BBS/E/C/000I03320 and BBS/E/C/000I0330), the National Natural Science Foundation of China (NO. 41571130083 and NO. 41725006) and the National Key Research and Development Program of China (No. 2016YFC0501601). All of the analyses and mapping were undertaken in R 3.5.3, the open source statistical software using the GWmodel package, v2.0-8 (Lu et al. 2014; Gollini et al. 2015). The code and data used in this paper can be downloaded from https://github.com/lexcomber/GWRroutemap.

# Appendix: Generic considerations and further guidance

## A.1 Characteristics and properties of the study data

### *Exploratory Data Analysis*

As in any statistical study, before any formal analysis is undertaken an exploratory data analysis or EDA is useful. The EDA should, at the bare minimum, consist of summary statistics, histograms, examination of correlations and the linear regression fit, together with specific spatial investigations described below. The EDA will confirm if worthwhile data relationships are present through the correlation and regression analysis, at least globally. It will also determine the presence of any problems with the data that need to be flagged or addressed. Common problems to address include that of non-linearity and outliers, where a data transformation (say, a log or square root) may be required, which may also provide a first step to dealing with error heteroskedacity. Global issues of predictor variable collinearity can also be addressed at this stage. In this respect, alternatives to the OLS-estimated linear regression fit, may be presented, such as a robust regression to deal with outlying relationships, a WLS fit to deal with error heteroskedacity and a ridge regression to deal with collinearity (see section 4 of the main text).

In addition to the EDA described, and in the context of GWR, the following considerations should also be investigated: (i) predictor variable specification, (ii) the presence of spatial predictors, (iii) evidence of spatial pattern in the response and predictors, (iv) effects of data pre-processing, (v) effects of sample size, and (vi) effects of sample configuration. These additional investigations are also exploratory and should be undertaken with the aim of understanding the data and to identify any characteristics that ultimately may affect a subsequent GWR analysis.

### *Predictor variable specification*

The first and most important consideration is to establish that there is some kind of expected relationship or process linking the response and predictor variables. That is, to confirm that the data have been collected with attributes that reflect either an underpinning research understanding of the problem, or with the aim of investigating the problem and to support the development of new understandings. So, a key question is whether *all* the required predictors are present and whether any spatial heterogeneity observed through a GWR analysis may simply be a consequence of missing predictors (i.e. global, linear model *misspecification*). This line of thought or even objection to GWR has been present from the outset (Brunsdon et al. 1998), and the same responses still apply: (a) the exploratory nature of GWR means its outputs can potentially guide the analyst to improve specification; (b) missing predictors may not be easily measured (e.g. too costly), and (c) the process is intrinsically spatial and local, where 'global truths' and 'stylised facts' are unlikely. Observe, the greater the number of predictors, the more likely it is that a linear regression (or a SAM) will identify some global truth and also that GWR may sometimes identify a spatial pattern when none actually exists (Paez et al. 2011; Harris 2019).

### *Spatial predictor variables*

A second consideration is whether *spatial* predictors are present? That is, predictors that are inherently spatial in nature. Useful ones such as the coordinates or region indicators can be more simply used in a linear regression or a SAM, as well as distance-based measures. Examples of the latter include distance to the city centre in an accessibility study, distance to fast food outlets as a variable in an obesity study or distance to sea in a lake acidification study. Such predictors should be avoided in any GWR study, as GWR itself employs distance-based analyses, so introducing distance related attributes can confound GWR results.

*Spatial pattern and autocorrelation amongst response and predictor variables*

At the exploratory stage of a GWR analysis, it is not only important to assess any residual autocorrelation from linear regression (and GWR) fits (see main text); but it is also important to investigate for spatial autocorrelation and co-autocorrelation in the response and predictors. This can be done simply through a series of Moran's *I* (and bivariate extensions of) analyses or can be done more thoroughly via the calculation and modelling of variograms and cross-variograms (e.g. Goovaerts 2001). Variographic assessments are particularly pertinent in that: (1) modelled variogram and cross-variogram ranges can help guide fixed bandwidth choice in GWR and (2) strong spatial co-autocorrelation amongst the variables can confound identification issues when choosing between a regression accounting for spatial heterogeneity effects (i.e. GWR) and a regression accounting for spatial autocorrelation effects (i.e. SAM) (Murakami et al. 2017; Geniaux and Martinetti 2018; Harris 2019).

Mapping the response and predictor variables is also key to determine if there is some spatial pattern and complements the more formal assessments, above. If spatial patterns or spatial dependencies are absent, then any spatial regression analysis (with a GWR or a SAM) should not be preferred over a non-spatial analysis with linear regression. In some instances, even a linear regression will hold no value, as all processes are purely random in nature with no linkages between them.

*Effects of data pre-processing*

Care must be taken when pre-processing the response and predictors prior to a GWR analysis. An analysis with raw data will commonly provide quite different outputs to that found using standardized and / or transformed data. This is somewhat highlighted in that MS-GWR has a thorny, and as of yet, unresolved calibration issue with the respect to standardizing the data or not, as a different set of bandwidth estimates will result (Oshan et al. 2019; Lu et al. 2019). In this study, the bandwidths for the MS-GWR models (of section 3 in the main text) were first estimated using centred data, which also provided computational savings. These bandwidths were then pre-specified in a second MS-GWR calibration, but now with the raw data, so that the MS-GWR coefficient sets could be directly compared to those found from the linear regression, SAM, standard GWR and MX-GWR.

*Sample size*

Ultimately, any evaluation of whether there are sufficient records for a GWR analysis depends on the nature of the spatial process being investigated, where small data sets can suffice if the process is well-behaved (i.e. relationships are expected to vary smoothly, the data has no *secondary* issues, there are relatively few predictors, etc.). Conversely, a GWR analysis with a large data set may still suffer from insufficient information if the spatial process is not well-behaved requiring a detailed and complex GWR route map. Páez et al. (2011) suggested a minimum of $n = 160$ records are appropriate for a GWR analysis, although this should never be considered a rule, only a loose guide.

For massive data sets (say, $n >> 10,000$), computational problems can arise, particularly in respect of any automated bandwidth selection procedure, where the computationally demanding back-fitting calibrations for MX-GWR and MS-GWR can be prohibitive to their use in *Big Data* studies. Computational burden can be alleviated through some combination of the following: (a) the judicious use of small but spatially-representative data subsets for bandwidth selection, (b) the use of centred predictor data (Lu et al. 2019, see above); (c) parallelisation (Harris et al. 2010c; Tran et al. 2016; Li et al. 2019), (d) the use of low level coding (e.g. C++ in 'GWmodel'), and (e) the pre-compression of GWR's matrices and vectors in scalable GWR (Murakami et al. 2019), also in 'GWmodel'.

***Sample configuration***

Other than the recommendation to specify adaptive bandwidths when the sample configuration is highly uneven in layout, little research has been conducted on the consequences of different sample configurations on a GWR analysis (aside from that given in Ye et al. 2017 in the context of prediction). For point support studies, such as those in soil science, it is likely that sample configurations recommended in geostatistical methodology (e.g. Webster and Lark 2012) are transferable to GWR. For example, the use of a regular or random stratified sampling grid, say. As in geostatistical studies, if the sampling is too coarse, processes at a finer scale will go unnoticed (i.e. small-scale spatial dependencies with kriging or highly localised spatial relationships with GWR). In the extreme, all spatial effects can go unnoticed resulting in choosing a linear regression fit simply due to poor sample design. Equally as important are the biasing effects due to preferential sampling where areas of perceived interest (e.g. high levels of soil contamination) are sampled more intensively than others. Here the preferentially sampled data require down-weighting or declustering in some manner to avoid (potentially severe) bias in GWR bandwidth estimation, model fit and coefficient estimation. It is likely that the geostatistical declustering procedures outlined in Diggle et al. (2010) are broadly transferrable to a GWR analysis.

## A.2 Further influences on the geographical weights

As indicated, the weightings in GWR are determined by a kernel function, where its bandwidth can be of a fixed or an adaptive distance form. Thus, experimentation with different kernel types (e.g. Gollini et al. 2015) and different bandwidth forms (see Table 4 of the main text) will directly influence GWR's weights and potentially the interpretation of its outputs.

In this respect, experimentation with both a discontinuous (e.g. box-car, bi-square, tri-cube) and a continuous (e.g. Gaussian, Exponential) kernel is recommended as it can provide clarity to any spatially-varying relationships observed. A box-car kernel (i.e. GWR defaulting to a moving-window regression) is useful in that it can return the corresponding global regression when a 100% adaptive bandwidth is specified. Furthermore, its highly discontinuous nature can be useful in the detection of outliers (Lloyd and Shuttleworth 2005). A continuous kernel is useful (and may be the only viable option) when sample size is small (say, $n < 100$) as it ensures that all data influence each local regression fit, yielding a certain robustness that is not possible when a discontinuous kernel is specified, as it can only use data subsets for each local regression fit. Adaptive bandwidths for discontinuous distance-decay kernels set at above 100% would address this, but this option is rarely available in GWR software packages.

Other GWR specifications are possible that influence its weights and can be worthy of investigation depending on the diversity of the sample data and the complexity of the geography. This includes the use of: (a) non-Euclidean distance metrics in standard GWR, MX-GWR and MS-GWR (Lu et al. 2014; 2015; 2017; 2018; 2019; Comber et al. 2018a) (e.g. for process along some urban transportation or river network), (b) double weighting schemes in contextualized GWR for hierarchical processes (Harris et al. 2013), (c) double weighting schemes in robust GWR (see Fotheringham et al. 2002; Harris et al. 2010a), (d) GWR with location-specific bandwidths (Páez et al. 2002a; b; Comber et al. 2018b), (e) locally linear GWR, which can improve fit and reduce coefficient bias over standard GWR (Wang et al. 2008; Páez et al. 2011; Zhang and Mei 2011), and (f) anisotropic GWR where weights decay at different rates according to directional relationships (Páez 2004).

## A.3 Inference options in GWR

Inference in GWR is somewhat compromised by there being no-one single model, but a collection of models re-using sample data at multiple locations. This entails that a valid probability model is unavailable with GWR, making inference biased and problematic. In this respect, Bayesian SVC models have a distinct advantage as they provide a valid and richer inferential framework for testing hypotheses (Gelfand et al. 2003; 2004; Finley 2011), but relative to GWR, can suffer analytically and computationally making them unusable in certain *Big Data* situations.

In section 3 of the main text, local inference directly used the local coefficients and their standard errors in an analogous way to that routinely done with the linear regression. This rudimentary approach has been referred to as *pseudo t-tests*, reflecting the caveats above (e.g. Harris et al. 2010a), but can provide cautiously reasonable results (Harris 2019). Improvements (adjustments) to this approach are provided in da Silva and Fotheringham (2016) regarding the inherent multiple hypothesis testing issue, which has also been extended to MS-GWR (Yu et al. 2019). Local inference in GWR can also be improved via the use of local bootstrap tests (Harris et al. 2017). Local inference can test whether coefficients significantly differ to zero or significantly differ to the same coefficient estimated globally through some fixed coefficient model (Harris et al. 2017; Harris 2019).

Local tests provide mappable outputs, but it is also possible to conduct tests for coefficient nonstationary against a fixed coefficient null hypothesis for each relationship of the regression model. For example, Nakaya et al. (2005) examined the variability of GWR coefficients by comparing standard with mixed models (all in a generalized form). For example, Nakaya (2015) added a deviance-based test for generalized GWR models. Similarly, Harris et al. (2017) proposed a parametric bootstrap test to compare coefficient estimates from standard GWR to those from a linear regression and SAMs, while Mei et al. (2016) proposed a non-parametric bootstrap test to compare coefficient estimates from standard GWR with those from MX-GWR. These approaches are generic and could be easily extended to all GWR forms.

## A.4 GWR as a spatial predictor

If the aim is spatial prediction at un-sampled locations, then almost all forms of GWR can be used, some of which have: (i) been specifically designed for this use purpose (e.g. Harris et al. 2010b; 2011), (ii) hybridised with kriging (Harris et al. 2010b; Harris and Juggins 2011; Kumar et al. 2012; Robinson et al. 2013; Zeng et al. 2016; Guo et al. 2017; Ye et al. 2017; Chen et al. 2019), (iii) designed to predict on specific supports (Lin et al. 2011; Jin et al. 2018), and (iv) re-purposed GWR within a GAM framework and fitted using penalized splines (Nogués-Bravo 2009). However, given a plethora of alternative prediction models exist within Geostatistical (Cressie 1993), Geographical (Haining 2003) and Machine Learning (Li et al. 2011) paradigms, results from a GWR-based predictor should always be compared with alternatives for objective context (Páez et al. 2008; Lloyd 2010; Harris et al. 2010b; Harris et al. 2011; Harris and Juggins 2011; Monteys et al. 2015; Song et al. 2016).

Alternative prediction models can also have similar non-stationary relationship options. For example, when the classic kriging with an external drift (KED) model (e.g. Chiles and Delfiner 1999) is specified with local kriging neighbourhoods rather than a single, unique one (Harris et al.

2010b; 2011; Monteys et al. 2015). A further consideration, one that is often over-looked, is that of prediction uncertainty. For GWR, such estimates can be found through: (a) a standard GWR predictor (Leung et al. 2000), (b) a GWR kriging hybrid (Harris et al. 2010b), (c) a heteroskedastic GWR predictor (Harris et al. 2011) and (d) a GWR indicator kriging hybrid (Harris and Juggins 2011). Bayesian SVC models, but now calibrated for prediction, can provide a superior inferential framework for prediction uncertainty to that based on GWR (Finley 2011).

## A.5 GWR development through simulation experiments

Finally, many developments of GWR have utilized simulation experiments to objectively demonstrate the value of a newly proposed GWR model or to demonstrate the value of an existing GWR model in relation to an alternative SVC model. These simulation experiment generate regression coefficient processes with known spatial characteristics. Different simulation designs exist, where those worthy of following include that of Wang et al. (2008), Wheeler and Calder (2007), Wheeler (2009), Finley (2011), Harris et al. (2017), Oshan and Fotheringham (2018), Wolf et al. (2018), Murakami et al. (2019) and Harris (2019).